\documentclass[12pt]{amsart}

\usepackage{amsmath,amssymb,graphicx,multicol}
\usepackage{color}
\usepackage[margin = 2cm]{geometry}
\usepackage{caption}

\newcommand{\ve}{\varepsilon}

\title{Minimal model of directed cell motility on patterned substrates}
\author{Matthew S. Mizuhara}
\thanks{MSM: Department of Mathematics, The Pennsylvania State University, University Park, PA 16802, USA. The work of MSM was supported by the Department of Defense (DoD) through the National Defense Science \& Engineering Graduate Fellowship (NDSEG) Program and by NSF grant DMS-1405769.} 
\email{msm344@psu.edu}
\author{Leonid Berlyand}
\thanks{LB: Department of Mathematics, The Pennsylvania State University, University Park, PA 16802, USA. The work of LB was supported by NSF grant DMS-1405769.}
\email{lvb2@psu.edu}
\author{Igor S. Aronson}
\thanks{ISA: Departments of Biomedical Engineering, Chemistry and Mathematics, The Pennsylvania State University, University Park, PA 16802, USA. The research of I.S.A. was supported by the US Department of Energy (DOE), Office of Science, Basic Energy Sciences (BES), Materials Science and Engineering Division.}
\email{isa12@psu.edu}

\begin{document}

    \maketitle
    \begin{abstract}
      Crawling cell motility is vital to many biological processes such as wound healing and the immune response. Using a minimal model we investigate the effects of patterned substrate adhesiveness and biophysical cell parameters on the direction of cell motion. We show that cells with low adhesion site formation rates may move perpendicular to adhesive stripes while a those with high adhesion site formation rates results in motility only parallel to the substrate stripes. We explore the effects of varying the substrate pattern and the strength of actin polymerization on the directionality of the crawling cell; these results have applications in motile cell sorting and guiding on engineered substrates.
      
 \end{abstract}

\bigskip


\section{Introduction}

Eukaryotic cell motility is crucial to many biological processes ranging from wound healing \cite{Moh03} and the immune response \cite{BanSte98} to cancer metastasis \cite{BirBir03}. The underlying biophysical mechanisms leading to persistent cell motion are generally understood: actin treadmilling (polymerization) drives protrusions at the cell front \cite{Pol03,PolBor03} while adhesion complexes transfer traction forces to the substrate \cite{Bar11} and myosin motors produce contractions at the cell rear \cite{Abe80,SheFelGal98}. However the interactions of these mechanisms with external stimuli, e.g., varying substrate properties, remains relatively unexplored. 


Prototypical cells for experiments, and subsequently, mathematical models, are keratocytes (e.g., harvested from fish scales). In homogeneous environments, once individual keratocytes initiate motion they exhibit characteristic crescent profiles and maintain essentially constant shape, speed, and direction \cite{Ker08,VerSviBor99}. Moreover as the characteristic cell length/width is two orders of magnitude larger than the height while motile, keratocytes are amenable to 2D models and thus may be considered the simplest cells for development of mathematical models (for a review on advances in 3D modeling techniques see, e.g., \cite{MogOdd11}).

For a general overview of both biological aspects of actin driven cell motility and of several modeling approaches we recommend the survey \cite{Mog09}.  In particular, in recent years both free boundary and phase-field models have been extremely successful in replicating, explaining, and predicting experimental results (see, e.g., \cite{CamZhaLi17,ziebert2011model,RecPutTru13,RubJacMog05,Sha10,Tjh15,lober2015collisions}), see for review \cite{ziebert2016computational}. We highlight that recent mathematical analyses have elucidated biological mechanisms \cite{BerPotRyb16,MizBerRybZha16, MizZha17,RecTru16}, and numerical simulations have described a wide range of behaviors ranging from motility initiation via stochastic fluctuations \cite{BarLeeAllTheMog15} to capturing various modes of motility such as stick-slip and bipedal motions \cite{BarAllJulThe10,LobZieAra14,Zie13}.  

The focus of our study is the effect of variable substrate properties on the direction of the cell's motion. In experimental settings the substrate may be coated non-homogeneously with fibronectins which allow for the creation of adhesion complexes, resulting in regions of variable adhesiveness. In \cite{CsuQuiDan07} microcontact printing of regions enriched with (or depleted of) fibronectins resulted in alternating substrate stripes of high and low adhesiveness with period smaller than the cell size. In this setting the authors of \cite{CsuQuiDan07} found that keratocytes exhibited directed motion parallel to the adhesive stripes.

We briefly mention other works which study the effect of non-homogeneous substrates. If the cell's size is smaller than the width of the adhesive stripe then both experiment and modeling \cite{RolNakYam12,Zie13} observe that the cell may be contained within adhesive regions: when the cell encounters a region of low adhesiveness it will change its direction in order to remain on regions of high adhesiveness. Moreover, it has been observed that cell morphologies can be controlled when placed on specific adhesive geometries \cite{CheMrkHuaWhiIng98} and directionality can be controlled on substrates with varying stiffness \cite{LoHonDemWan00,TriLeD12}. As such, controlling and predicting the motility of keratocytes on engineered patterned substrates has direct applications to cell screening and sorting both in biomimetics \cite{SheGar11} and experimental settings.

%
%


In this work we study the directionality of a motile cell on a patterned substrate with alternating stripes of high and low adhesiveness; here the cell size is assumed to be larger than the stripe width so that the cell spans several regions of both high and low adhesiveness corresponding to the experimental setup of \cite{CsuQuiDan07}.
 We stress that the dynamics of adhesion site formation are, in general, complex \cite{BruGouPin94,BurChrl96,SacBru02,SmiSei05}. In \cite{Zie13} a phase-field model which included both dynamics of the adhesion site formation as well as substrate deformation (e.g., due to traction forces generated by the cell) was introduced. The authors introduced patterned adhesiveness and reproduced experimental results observed in \cite{CsuQuiDan07}. Surprisingly, the authors of \cite{Zie13} also obtained for certain parameter regimes that the cells move perpendicular (rather than parallel) to substrate adhesion stripes. More specifically, when the attachment rate of adhesion complexes is sufficiently high (e.g., when adhesive stripes had high adhesiveness), numerical simulations reproduce experimental results. When the attachment rate of adhesion complexes is low (e.g., when adhesive parts of stripes have low adhesiveness), numerical simulations show that cells move perpendicular to substrate stripes; this striking behavior has thus far remained unobserved in experiments and merits additional numerical investigation.


%

Due to the complexity of the phase-field model in \cite{Zie13} there is no clear, simple mechanism to explain and differentiate parallel/perpendicular motions.  To elucidate the effects of substrate adhesiveness patterns and biophysical parameters on the resultant direction of motion, we derive from the full system a minimal model of cell motility on patterned substrates which is capable of reproducing both parallel and perpendicular (to stripes) directed motions. This minimal model reveals the underlying mechanical processes which may give rise, in particular, to perpendicular motions. Moreover the reduction allows for efficient numerical experiments over a wide range of parameters revealing the dependence of direction of motion on biophysical (e.g., actin polymerization strength) and substrate properties (e.g., stripe sizes, adhesiveness). 

%
%





\section{Results}

\subsection{Description of the reduced model}

 In \cite{Zie13} a 2D phase-field model of cell motility was introduced to describe the onset and persistence of cell motility as well as a broad range of cell morphologies.  The model contains four coupled differential equations: a phase-field equation describing the location of the cell membrane, an evolving vector field representing the effect of the actin filament network, a scalar equation for density of adhesion sites, and a Kelvin-Voigt visco-elastic equation for the deformation of the substrate.

\begin{figure}
\centering
\includegraphics[width = .8\textwidth]{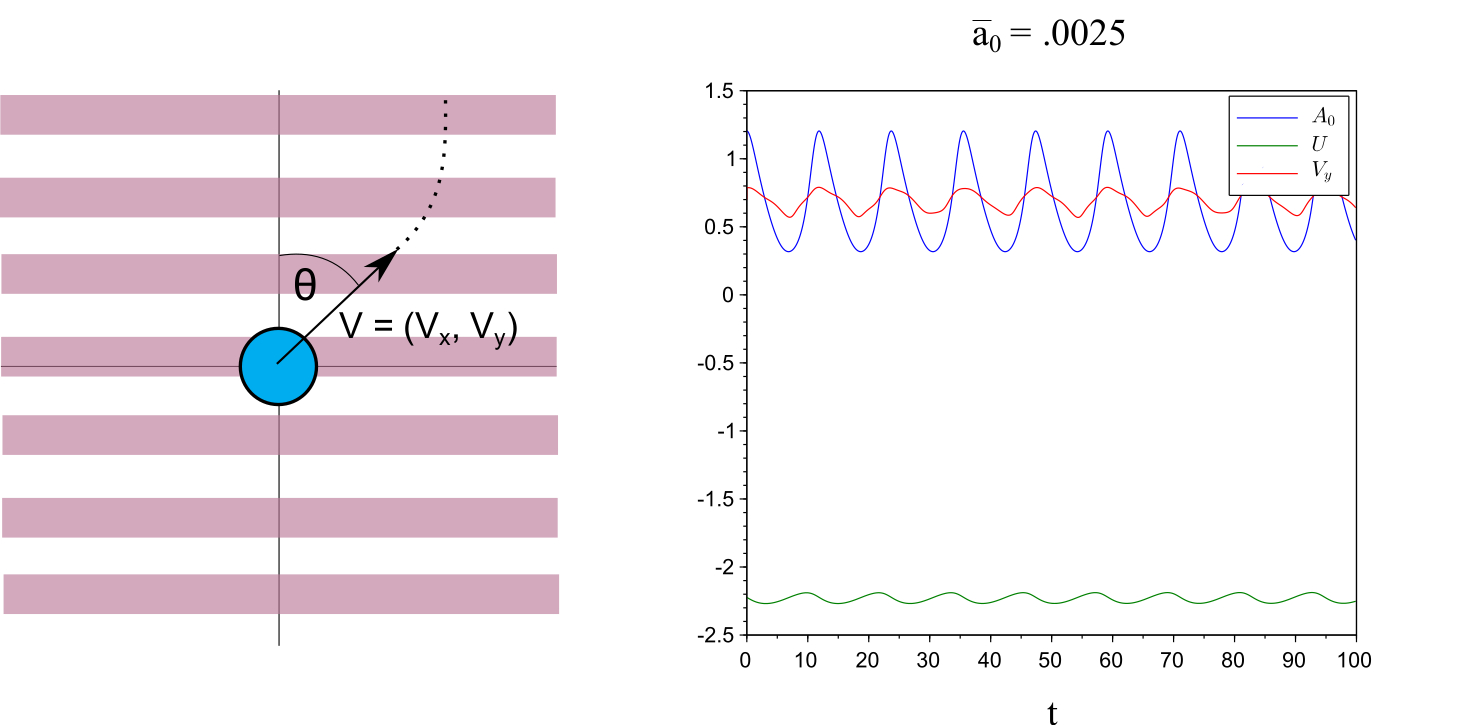}
\caption{(left) Sketch of cell motion on substrate with patterned stripes of adhesiveness described by \eqref{vxeqn}-\eqref{ueqn}; $\theta$ measures the deflection of the initial velocity from the vertical axis. (right) For small values of $\bar{a}_0$ the keratocyte experiences significant oscillations in velocity, adhesion density, and substrate deformations. Oscillations in number of adhesions and substrate deformations are linked to stick-slip motion. Taking $\bar{a}_0>.1$ results in convergence of $A_0$ and $U$ to equilibrium. Due to inhomogeneity of the substrate, the velocity $V_y$ still has oscillations although with smaller amplitude.}
\label{theta_picture}
\end{figure}

 As exploration of the full phase-field model is complicated we propose the following simplified system of differential and algebraic equations which track the location of the center of the cell $(x,y)$ as well as the effective adhesion of the cell to the substrate $A$ and effective substrate deformation $U$ (see Figure \ref{theta_picture} for an illustration of the model):

\begin{align}
\frac{d}{dt}x &= V_x = f(V_x,V_y,A,y), \label{vxeqn}\\
\frac{d}{dt}y &= V_y = g(V_x,V_y,A,y), \label{vyeqn}\\
A &= A_0 \left(\frac{1}{2}  (1+ \operatorname{sign}(\sin (k_0 y)))\right) \label{aeqn}\\
\frac{d}{dt}A_0 &= \bar{a}_0 - d(U) A_0 + \bar{a}_{nl}A_0^2-\bar{s}A_0^3 \label{a0eqn}\\
-\eta \frac{d}{dt} U &= GU + \bar{\alpha} A_0. \label{ueqn}
\end{align}

Here, $V_x, V_y$ are the center of mass velocities, $d(U):= \frac{1}{2}(1+\tanh[b(U^2-U_c^2)])$ is an effective cut-off function which destroys adhesions if the substrate deformation $U$ exceeds some critical value $U_c$, $k_0$ is  the wavenumber of microprinted stripe pattern.  Equation \eqref{aeqn} encodes the striped substrate by creating inhomogeneity in the $y$-direction.

System \eqref{vxeqn}-\eqref{ueqn} is derived from the phase-field model in \cite{Zie13} after several simplifying assumptions; a detailed derivation can be found in the Supplementary Materials. In this reduced model we aim to replicate both the numerically observed emergence of perpendicular motion to the stripes as well as the experimentally observed parallel motions whereby we may study the robustness of both as well as understand the biophysical mechanisms which give rise to each.

We note that the system \eqref{a0eqn}-\eqref{ueqn} is decoupled from equations \eqref{vxeqn}-\eqref{vyeqn} so that $A=A(t)$ enters \eqref{vxeqn}-\eqref{vyeqn} as a time varying coefficient depending on physical parameters. Since we will focus on the effect of varying initial conditions and physical parameters on the motion of the cell, the decoupling yields efficient numerical simulation. Indeed, we can compute $A_0$ and $U$ independently form equations for $V_x, V_y$, and thus  to reduce overall computation time and effectively decrease the search over initial conditions by two dimensions. See Section \ref{sec:methods} for a detailed description of the numerical methods used.

Although this decoupling is a vast simplification to the physical model, subsequent numerical study shows that we retain sufficient structure for meaningful results; future work may investigate the fully coupled system by replacing \eqref{ueqn} by the vector equation $-\eta \frac{d}{dt}{\bf U} = G{\bf U}+ {\bf V}$, where ${\bf V}=(V_x,V_y)$.

 \subsection{Smaller driving force required for motion on striped substrate} 
 
 We first consider the effect of a striped substrate on the minimal driving force required for persistent cell motion. As observed in \cite{Zie13} if we take $\bar{a}_0$ sufficiently small, e.g., $\bar{a}_0=.0025$, then the system \eqref{a0eqn}-\eqref{ueqn} tends to a limit cycle. On the other hand taking $\bar{a}_0$ large, say $\bar{a}_0 = .25$, \eqref{a0eqn}-\eqref{ueqn} tends to an equilibrium, suggesting the existence of a supercritical Hopf bifurcation in the parameter $\bar{a}_0$; the point of bifurcation is $\bar{a}_0\approx .1$. For simplicity we take $\bar{a}_0=.25$ so that the pair $(A_0,U)$ tends to an equilibrium $(A_\infty,U_\infty)$. Following \cite{Zie13}, we define $\kappa:=8\alpha\beta\tau_1^2 A_\infty/(81 R_0^2)$, the normalized driving force in the cell. In \cite{Zie13} a subcritical onset of motion was observed on homogeneous substrates: for $\kappa < \kappa_c \approx .746$ there is no persistent motion of the cell while for $\kappa>\kappa_c$ there is some finite minimal velocity of the cell. 
 
  Upon inclusion of substrate stripes the minimal driving force required for the onset of persistent motion decreases: $\kappa_c\approx .395$. 
  As expected, this motion is parallel to the substrate stripes. 
  This predicts that microcontact printing may allow for directed cell motion even for cells which cannot sustain persistent motion on homogeneous substrates. These results are analogous to observations in \cite{PeyKalCohRun12} where it was shown that the distance traveled by a motile cell is larger when restrictions to the cell geometry are imposed (e.g., diameter of the background matrix is made smaller than the cell diameter). Moreover results of \cite{RodSch13} show that micro-contact printing may dominate chemical cues for contraction driven motile cells (e.g., fibroblasts). 
 
\subsection{Robustness of vertical trajectories decreases as adhesion formation rate increases}\label{sec:stability}
 
 Since $f(0,V_y,A,y) = 0$ then $V_x = 0$ is always a trivial solution of $V_x = f(V_x,V_y,A,y)$. We thus first investigate the existence of persistent motion perpendicular to the substrate pattern in the reduced system \eqref{vyeqn}- \eqref{ueqn} with $V_x=0$.  Numerical analysis  shows that persistent motion perpendicular to stripes is possible over a large range of physical parameters. For subsequent results we assume to use parameter values as in Table \ref{table_params} unless otherwise mentioned.
 
\begin{table} 
     \[
     \begin{tabular}{|c|c|c|}
     \hline 
     parameter & value & description \\
     \hline 
     $k_0$ & 2.5 & wave number of substrate pattern \\
     $\bar{a}_0$ & $.0025 - .25$ & linear attachment rate of adhesion sites\\
     $\bar{a}_{nl}$ & 1.5 & effective collective (nonlinear) attachment rate of adhesion sites \\
     $\bar{s} $ & 1 & local saturation of adhesion sites\\
     $\eta$ & 10 & dissipation in the adhesive layer \\
      $G$ & .15 & substrate stiffness \\
    $\bar{\alpha}$ & $.5$ & effective propulsion strength  \\
     $b$ & 5 & sharpness of breaking function \\
     $U_c$ & $\sqrt{5}$ & critical extension to break adhesive contacts \\
     \hline
     \end{tabular} 
     \]
     \caption{Physical parameter values used in numerical simulations.} 
       \label{table_params}
     \end{table}

  Surprisingly our numerical analysis also shows persistent vertical trajectories exist for all values of $\bar{a}_0$ (in contrast to results obtained in the full PDE model). First taking the effective linear attachment rate $\bar{a}_0 = .0025$, the $A_0$-$U$ system exhibits a limit cycle (i.e., stick-slip behavior), which in turn results in large oscillations in the $y$ velocity, see Figure \ref{theta_picture}. Taking $\bar{a}_0 =.25$, then the $A_0$-$U$ system tends to equilibrium but a purely perpendicular motion still exists. Oscillations in the $y$ velocity result only from the non-homogeneity of the substrate and thus are smaller than in the case of oscillating $A_0$ and $U_0$. 
Since persistent motion perpendicular to stripes is observed for all values of $\bar{a}_0$, in order to corroborate our results with the full PDE simulations, we aim to understand the robustness of perpendicular trajectories via a stability analysis of the vertical motion as a function of $\bar{a}_0$. That is, we consider the long time direction of motion if the initial $x$-velocity $V_x$ is chosen to be non-zero.  Due to the algebraic dependence of $V_x$ in \eqref{vxeqn} classical linear stability analysis techniques are not available. 
Thus, to study stability of vertical trajectories, we exhaustively search initial conditions numerically in order to see long time asymptotics,
for details see Section \ref{sec:methods}.

\begin{figure}
\includegraphics[width = .45\textwidth]{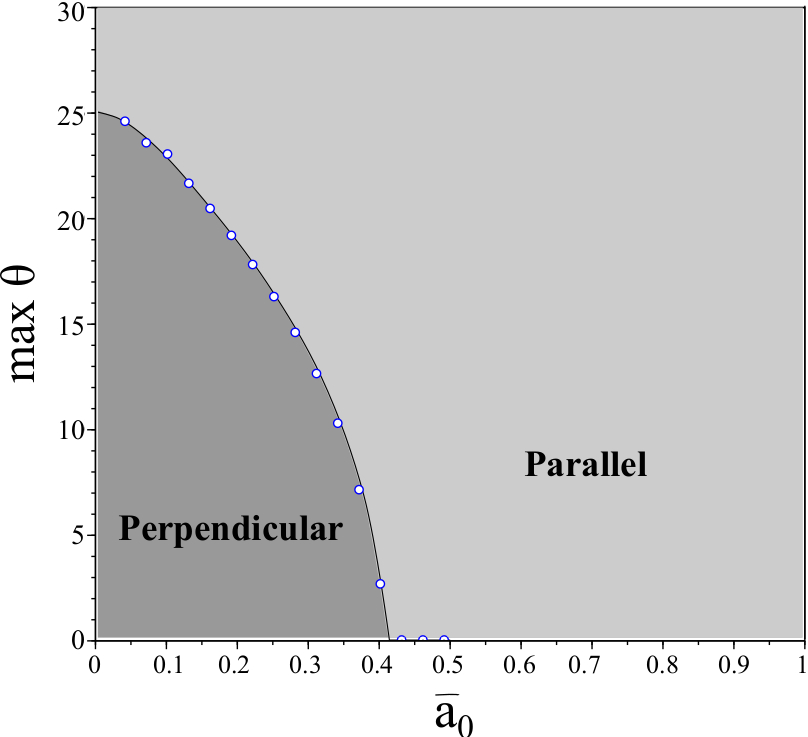}
\includegraphics[width=.45\textwidth]{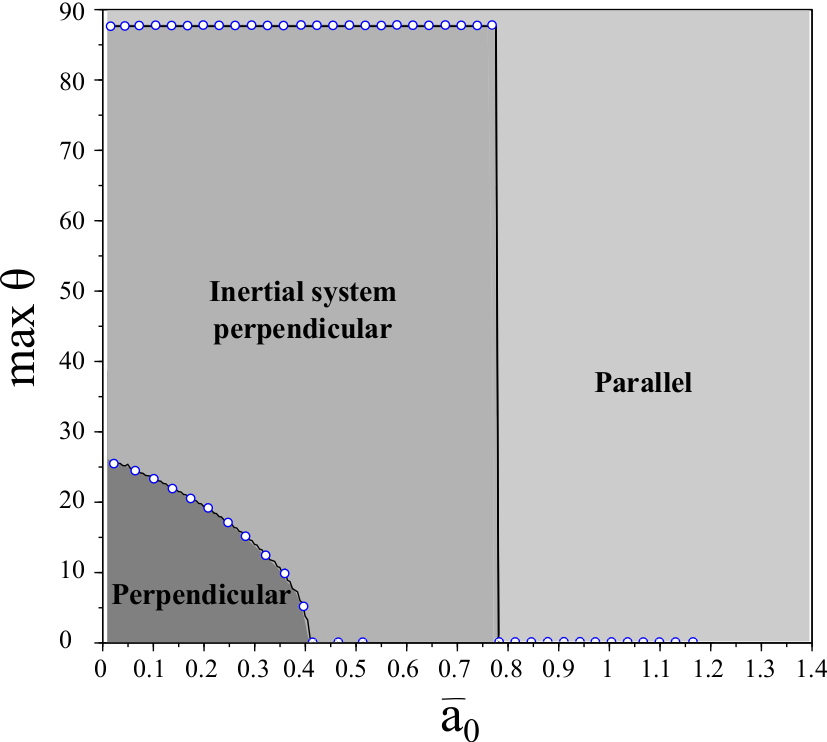}
\caption{(left) The maximal angle of the initial velocity for which the cell tends to persistent perpendicular motion is $\max \theta$. For small values of $\bar{a}_0$ perpendicular motion is possible ($\max\theta >0$). For large values of $\bar{a}_0$ only parallel motion is possible. Simulations are done using the system  \eqref{vxeqn}-\eqref{ueqn}. (right) Simulation of \eqref{eq3}-\eqref{eq4} with \eqref{aeqn}-\eqref{ueqn} shows a larger basin of attraction for perpendicular motion. Even for nearly horizontal initial conditions the cell may eventually move perpendicular to stripes, as originally observed in the full phase-field model \cite{Zie13}. Dependence of $\max\theta$ on $\bar{a}_0$ is no longer continuous.}
\label{fig:inertia_phase}
\end{figure}

We define $\theta$ to be the angle (in degrees) of the initial velocity of the cell from the positive vertical axis, see Figure \ref{theta_picture}. For each $\bar{a}_0$ we exhaustively search initial conditions $y_0$ in order to determine the maximal value of $\theta$ which gives rise to persistent vertical motion. We define $\max \theta$ to be the maximal $\theta$ over all possible initial conditions. The dependence of $\max\theta$ on $\bar{a}_0$ is shown in Figure \ref{fig:inertia_phase}. 

We observe that $\max\theta$ is largest where $a_0$ is small, indicating that vertical motion is more robust in this regime.  This corresponds with the results in \cite{Zie13}; in particular $\max\theta$ is largest in the regime that the cell undergoes stick-slip motion (i.e., limit cycles in the $A_0-U$ system).

When $\bar{a}_0$ is sufficiently large, the pair $(A_0,U)$ tends to an equilibrium $(A_\infty,U_\infty)$. The equilibria $A_\infty$ and $U_\infty$ both increase with $\bar{a}_0$. We observe that  $(A_\infty,U_\infty)$ remains relatively constant for all $\bar{a}_0 < .3$ (e.g., $.65\leq A_\infty\leq .7$). However, $(A_\infty,U_\infty)$ begins to very rapidly increase for $\bar{a}_0>.3$ (e.g., $A_\infty \approx .88$ for $\bar{a}_0=.4$). Interestingly $U_\infty$ is approximately $U_\infty \approx R_0$ when $\max\theta=0$.  Although substrate variations of the order of magnitude of the cell size are unphysical (an artifact of our approximations), we conclude that large substrate variations may lead to stabilization of parallel motion whereas small substrate variations allow for perpendicular motion to stripes. This hypothesis agrees with previous numerical and experimental evidence which shows that cells may overcome variations in substrate stiffness provided the substrate is sufficiently stiff \cite{Zie13,RolNakYam12}. 




\subsection{Aspect ratio of substrate stripes sorts cells depending on actin polymerization strength}

We investigate the motion of cells on substrates with striped patterns of adhesiveness where the ratio of adhesive stripe width to non-adhesive stripe width is not necessarily equal. Let $L_1$ be the width of the adhesive stripe and $L_2$ to be the width of the non-adhesive stripe so that $L:=L_1+L_2$ is the period of the substrate pattern. We investigate the effect of varying the ratio $L_1/L$ on $\max\theta$. 

\begin{figure}
\includegraphics[width = \textwidth]{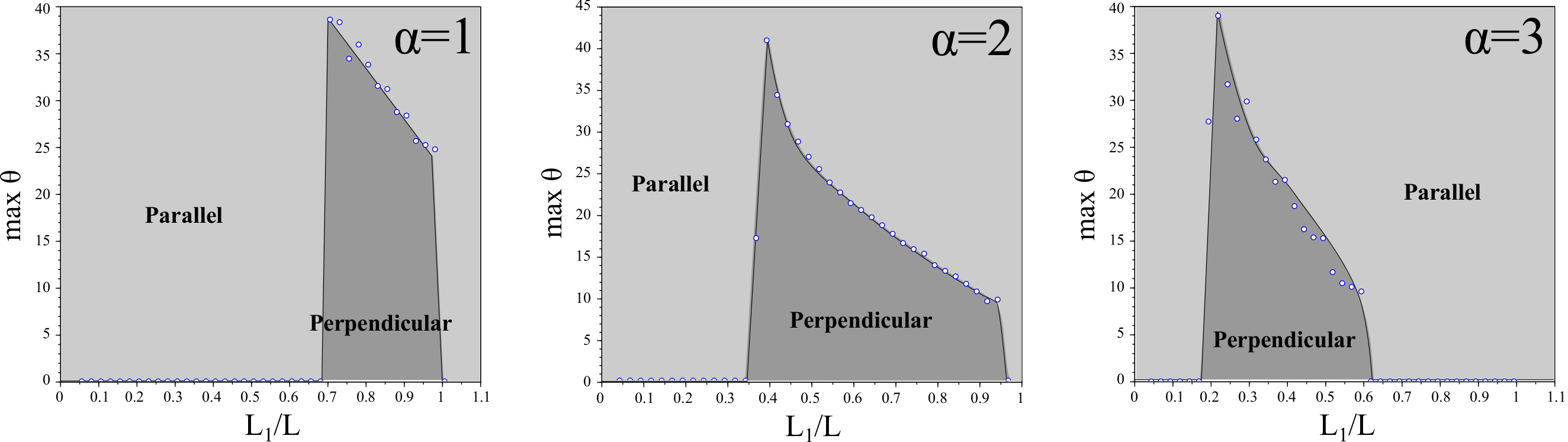}
\caption{The value of $\alpha$ and substrate pattern affect the direction of keratocyte motion. The value of $\alpha$ measures the rate of advection of the cell by the actin network and is a function, e.g., of the integrin ligand bond strength and actin filament stiffness. The patterned substrate has constant period $L$ but has varying width $L_1$ of the adhesive stripe. It is seen that if $\alpha$ is low then the cell requires a wider adhesive stripe in order to move perpendicular to the stripe and vice versa. This suggests a mechanism for cell guiding and sorting.}
\label{fig:effect_of_alpha}
\end{figure}

Here, we keep $\bar{a}_0$ constant ($\bar{a}_0 = .0025$) and vary $L_1$. Since $\alpha$ is a key physical parameter measuring the strength of actin polymerization, we additionally investigate how changing $\alpha$ changes these data. The results are summarized in Figure \ref{fig:effect_of_alpha}.

It is clear that if $L_1=0$ or $L_1=L$ then the cell cannot create a biased directionality and $\max\theta =0$.  This is expected since if the adhesive stripe is too small then the cell cannot develop a sufficient number of adhesive bonds to the substrate to initiate persistent motion. On the other hand if $L\approx 1$ the substrate is entirely adhesive and the substrate is effectively homogeneous.


%
%


Interestingly we observe a monotonic dependence between the actin polymerization strength $\alpha$ and the percentage of adhesiveness of the substrate which results in perpendicular motion to stripes. If the substrate is predominantly non-adhesive then the cell requires high actin polymerization strength to generate perpendicular motion and vice-versa. This suggests that different effectiveness of internal biophysical parameters may lead to different behaviors of the cell depending on the shape of the patterned substrate, providing evidence for cell sorting and directed cell motility. 

\subsection{Memory increases the basin of attraction of persistent perpendicular motion}

Thus far, the basin of attraction for vertical motion has been bounded by $\max\theta < 40$. However in full PDE simulations even with motion initialized parallel to the stripes, the cell switches directions and begin moving perpendicular to the stripes (provided $a_0$ is sufficiently small) so that $\max\theta \approx 90$.
We predict that this discrepancy arises from the lack of memory  in the reduction \eqref{vxeqn} - \eqref{ueqn}. That is, we neglect both shape deformation and directional memory (or inertia) which are present in the full PDE model. In particular, there is finite time relaxation of the actin vector field which is neglected in the reduced equations.

To include these memory effects we modify \eqref{vxeqn}-\eqref{vyeqn}:
\begin{equation}\label{eq3}
\varepsilon \ddot{x} + \dot{x} = f(\dot{x},\dot{y},A,y)
\end{equation}
\begin{equation}\label{eq4}
\varepsilon \ddot{y} + \dot{y} = f(\dot{x},\dot{y},A,y),
\end{equation}
where $\varepsilon$ is an effective memory (or )inertial) coefficient characterizing the time of relaxation of the cell's directional inertia. This coefficient is a function of both shape deformations and relaxation of the actin polymerization field. Mathematically it is also important to note that $\varepsilon >0$ provides regularization so that \eqref{eq3}-\eqref{eq4} absent in \eqref{vxeqn}-\eqref{vyeqn}. We compare numerical results with those results obtained for $\varepsilon=0$, see Figure \ref{fig:inertia_phase}. 
 Again, our interest is to study the dependence on $\max\theta$ as a function of $\bar{a}_0$. 




%

As \eqref{eq3}-\eqref{eq4} is a singular perturbation of \eqref{vxeqn}-\eqref{vyeqn}, it is natural that the behavior drastically changes qualitatively.  However, it is surprising that the value of $\max\theta$ becomes approximately piecewise constant in numerical studies. Moreover the maximum value of $\bar{a}_0$ for which $\max\theta>0$ decreases as $\varepsilon$ increases: a cell with longer memory is less likely to overcome the nonadhesive stripes and exhibit perpendicular motion. However we observe that even for $\ve=.005$ the positive values of $\max\theta$ are very large: $\max\theta \approx 77$, agreeing with simulations in \cite{Zie13}. 
This suggests that memory 
of shape deformations as well as persistence of the actin network may be correlated with persistent motion perpendicular to the stripes.

\section{Conclusions}

We have presented a reduced system derived from the full phase-field model presented in \cite{Zie13}. The reduced system is much less computationally expensive to simulate while still retaining the crucial solution behaviors. We reproduce both perpendicular and parallel persistent cell motions on periodically striped substrates.  Our numerical study  indicates that the robustness of vertical and horizontal motions can be quantified via consideration of long time asymptotics over all initial conditions (measured by $\max\theta$). These simulations indicate that low adhesion site formation rates are necessary for persistence of perpendicular motions. Physically we expect that if the number of adhesion sites is too high then the forces ``pulling'' the cell back towards the adhesive part of the substrate will prevent the cell from overcoming the arrest caused by the nonadhesive part of the substrate. 

 Moreover, we exhibited a inverse, monotonic dependence of actin polymerization strength on the amount of adhesive substrate stripes allowing for perpendicular motion. We predict that carefully engineered substrates could potentially use this correlation in applications of cell sorting and directed cell motility.
 
 Finally, by including memory (e.g. due to finite cell deformation relaxation time)   we obtain a singularly perturbed system which shows better agreement with full phase-field simulations: for small values of $\bar{a}_0$ the initial velocity may be almost parallel to stripes and still we observe persistent perpendicular motion over long time.  These results suggest that shape deformation and directional cell memory  correlate with the ability of the cell to move perpendicular to the substrate stripes. We believe that this methodology of reduction of a full PDE model to finite dimensional models (e.g., systems of ordinary differential equations) may be applicable to a wide range of physical models and may provide new insights via both analytical and numerical study of the resultant, simplified systems.

\section{Methods}\label{sec:methods}
We solve system \eqref{vxeqn} - \eqref{ueqn} using a forward finite difference scheme. In particular $V_x$ is defined via an algebraic equation, as opposed to a differential equation. As such to solve the implicitly defined velocities $V_x$ and $V_y$ we require a nonlinear function solver. We use a predefined function in Scilab (fsolve), which is based on an iterative Powell hybrid method, initializing the solver at the solution of the previous time step.

In order to conduct an exhaustive search through all initial conditions we first simulate the longtime dynamics of \eqref{a0eqn}-\eqref{ueqn} to determine either the value of the equilibrium point or the stable limit cycle. In the former case, we assign the limiting equilibrium value as the initial conditions for $A_0$ and $U$.  In the latter case, we may fix any point on the limit cycle (for consistency, we choose the point on the limit cycle where $A_0$ is maximal). 

When viewed as an algebraic system, \eqref{vxeqn}-\eqref{vyeqn} has non-unique solutions $(V_x,V_y)$ for fixed $y$. Thus for a fixed value of $y_0 \in [0, k_0/2\pi]$ we first compute all admissible initial velocities $V_x(0),V_y(0)$ and simulate the long time dynamics of all to deduce if either (i) the motion becomes eventually horizontal, (ii) the motion becomes eventually vertical, or (iii) there is no continuous in time velocities which solve \eqref{vxeqn}-\eqref{vyeqn} for all time. Since the equations \eqref{vxeqn}-\eqref{vyeqn} are nonlinearly coupled, case (iii) may arise, for example, as the result of bifurcations in the $V_x$-$V_y$ solution plane as $A$ or $y$ vary.  We recall that there are generically non-unique solutions to \eqref{vxeqn}-\eqref{vyeqn}, so if such a discontinuity occurs, it is deemed unphysical and so we omit this scenario from our analysis.

We additionally note that solving \eqref{eq3}-\eqref{eq4} requires small time steps $\Delta t\ll \varepsilon$ to ensure convergence as it is a singularly perturbed system. In general we may take any initial velocities $V_x(0),V_y(0)$, however we restrict ourselves to initial conditions which are compatible with initial velocities computed in the case that $\varepsilon=0$ so that results may be compared: given an initial $y_0$ we initialize $V_x(0),V_y(0)$ to be admissible solutions to the system \eqref{vxeqn}-\eqref{vyeqn}.

\section{Appendix}
 
 \subsection{Phase-field model}

 In \cite{Zie12} a 2D phase-field model of cell motility was introduced to describe the onset and persistence of cell motility as well as a broad range of cell morphologies.  This model contains two PDEs. First, a phase-field parameter $\rho$ describes the location of the cell membrane (i.e., $\rho \approx 1$ on the interior of the cell and $\rho \approx 0$ outside of the cell). Second, the motion of the cell membrane is actively driven by a vector field $p$ which models the averaged orientation field of the actin filament network. \\
 The coupling of these equations reflects two main experimental observations: (i) the nucleation of branches of actin filaments near the cell membrane via Wiskott-Aldrich syndrome proteins (WASP) activation of the ARP2/3 complex and (ii) presence of actin filaments allows for the creation of adhesive contacts and subsequent transfer of momentum of the polymerizing actin network to the substrate driving the cell membrane.
 Additionally the model enforces approximate volume preservation. The full PDE model introduced in \cite{Zie12} is:
 
 \begin{equation}\label{rhoeqn}
 \partial_t \rho = D_\rho \Delta \rho - (1-\rho)(\delta-\rho)\rho - \alpha Ap\cdot (\nabla \rho),
 \end{equation}
 \begin{equation}\label{peqn}
 \partial_t p = - \tau_1^{-1} p -\tau_2^{-1}(1-\rho^2)p-\beta \nabla \rho - \gamma [(\nabla \rho)\cdot p]p.
 \end{equation}
 For our subsequent analysis we assume that $\tau_2^{-1}=\gamma = 0$. In particular, the term containing $\gamma$ accounts for symmetry breaking due to myosin driven contraction in the rear of the cell; since our subsequent analysis initializes data with non-zero velocities, this symmetry breaking is not required for capturing persistent motion. Moreover it is mentioned in \cite{Zie13} that self-sustained cell motion is possible in the system \eqref{rhoeqn}-\eqref{peqn} even without this term. We refer the interested reader to \cite{Zie13,Zie12} for a full description of the model.
 
 
 In \cite{Zie12} it is assumed for simplicity that the friction generated from adhesion complexes is homogeneous so that $A$ is constant.  To account for the complex interaction between cell and substrate it is necessary to include the dynamics of adhesion site formation, see \cite{Zie13}:
 \begin{equation}\label{aeqn1}
 \partial_t A=D_A \Delta A+ \rho(a_0 p^2+a_{nl} A^2) - sA^3-d(U)A.
 \end{equation}
 These adhesion contacts describe integrin complexes which (through a series of intermediate proteins such as zyxin, talin, and vinculin) engage both the substrate and the cytoskeleton. 
 
 
 
 In the final term of \eqref{aeqn1}, the function $d(U)$ describes the coupling of $A$ to the average substrate deformation ${\bf U}(t)$, with $U=|{\bf U}|$:
 \begin{equation}\label{dueqn}
 d(U) = \frac{1}{2}(1+\tanh[b(U^2-U_c^2)]),
 \end{equation}
 where ${\bf U}$ satisfies
 \begin{equation}
 \frac{d}{dt}{\bf U} =-\frac{1}{\eta} (G {\bf U}+{\bf V}),
 \end{equation}
 where $\eta$ is the effective viscous friction of the substrate, $G$ is an effective spring constant and ${\bf V}$ is the velocity of the cell's center of mass. 
 That is, the substrate is viewed as a Kelvin-Voigt visco-elastic material and adhesion sites are broken if deformations exceed the threshold $U_c$. 
 
 As all attachment/detachment rates are effective parameters, they incorporate both characteristics of the adhesion complex as well as the substrate preparation.  Thus, spatial inhomogeneity in the substrate may be introduced through coefficients, e.g., spatial dependence of $a_0 = a_0(y)$.
 
 

 \subsection{Derivation of the reduced system}
 
 We reduce the system \eqref{rhoeqn} to a two-dimensional system for the location of the center of the cell $(x,y)$. All assumptions are analogous to those made in \cite{Zie13}. The biggest difference is that we do not assume that $V_y\ll 1$ and as such we have more complex coupling between all equations.  We first assume that the cell has fixed circular shape for all time:
 \begin{equation} 
 \rho(x,y,t) = \rho(x-x_0(t),y-y_0(t)),
 \label{rhoansatz}
 \end{equation}
 where
 \begin{equation}
 \rho(x,y) = \exp(-(x^2+y^2)/R_0^2).
 \end{equation}
 In particular, note that under the assumption \eqref{rhoansatz} we have $\delta \equiv \frac{1}{2}$. We denote the velocity of the center of the cell $V = (V_x,V_y) = (\dot{x}_0,\dot{y}_0)$.
 Multiplying \eqref{rhoeqn} by $\partial_x \rho$, and integrating over the domain yields
 \begin{equation}\label{vxeqn1}
 V_x \int (\partial_x \rho)^2 = \alpha \int A p \cdot (\nabla \rho)\partial_x \rho.
 \end{equation}
 Likewise multiplying \eqref{rhoeqn} by $\partial_y \rho$ and integrating we have
 \begin{equation}\label{vyeqn1}
 V_y \int (\partial_y \rho)^2 = \alpha \int A p \cdot (\nabla \rho)\partial_y \rho.
 \end{equation}
 We likewise assume that 
 \begin{equation}
 p(x,y,t) = p(x-x_0(t),y-y_0(t))
 \end{equation}
  to rewrite equation \eqref{peqn}:
 \begin{equation}
 -V \cdot \nabla p = -\tau_1^{-1} p -\beta \nabla \rho.
 \end{equation}
 Under the assumption that $|V|$ is small:
 \begin{align}
 p(x-\tau_1 V_x,y-\tau_1 V_y) &= p-\tau_1 V\cdot \nabla p + O(\tau_1^2V^2) \\
 &= -\tau_1 \beta \nabla \rho(x,y).
 \end{align}
 Thus to first order 
 \begin{equation}\label{newpeqn}
 p(x,y) = -\tau_1 \beta \nabla \rho(x+\tau_1 V_x, y+\tau_1 V_y).
 \end{equation}
 
 To simplify the equation for $A$ we first consider the problem on a homogeneous substrate (i.e., $a_0$ is constant) and consider the following ansatz for the adhesion sites density: $A = A_0(t) \rho(x,y,t)+\delta A_1(x,y,t)$ where $\delta$ is much smaller than the size of the cell. That is, we assume that $A$ is essentially spatially constant on the interior of the cell. Plugging this expansion into \eqref{aeqn1} and integrating over the domain we have to leading order:
 
 \begin{equation}
 \frac{d}{dt} A_0 = \bar{a}_0 -d(U)A_0+\bar{a}_{nl}A_0^2-\bar{s}A_0^3,
 \end{equation}
 where
 \begin{equation}\label{effective_coeff}
 \bar{a}_0 = a_0 \frac{\langle \rho p^2\rangle}{\langle \rho\rangle}, \bar{a}_{nl} = a_{nl} \frac{\langle \rho^3\rangle}{\langle \rho\rangle}, \bar{s} = s \frac{\langle \rho^3 \rangle}{\langle \rho \rangle}.
 \end{equation}
 Similarly we can simplify the equation for ${\bf U}$ to a scalar equation under the assumption that $U$ and $V$ are essentially co-linear. As in \cite{Zie13} we make the approximation $V(t) \approx \langle \alpha A_0 p\rangle \approx \bar \alpha A_0$, where $\bar{\alpha}$ is a numerical constant:
 \begin{equation}
 -\eta \partial_t U = GU+\bar{\alpha}A_0.
 \end{equation}
 Finally, in order to simulate the patterned substrate we approximate $A$ to have the form
 \begin{equation}\label{patterneda}
 A = A_0(t) \left(\frac{1}{2}  (1+ sign(\sin (k_0 y)))\right).
 \end{equation}
 For subsequent numerical simulations, we use the first two terms of the Fourier expansion of \eqref{patterneda} for simplicity and regularity.
 
 By plugging in \eqref{newpeqn}, \eqref{patterneda} into \eqref{vxeqn1}-\eqref{vyeqn1}, we derive the equations $V_x = f(V_x,V_y,A,y)$, $V_y = g(V_x,V_y,A,y)$, where $f$ and $g$ are defined
  { 
    \begin{align*}
    f &:= \left(\frac{4 \alpha A\beta}{243\pi R_0^4}\right)e^{-(2\tau_1^{2}(V_x^2+V_y^2))/(3R_0^2)}\tau_1^{2}V_x \cdot (3\pi(-3R_0^2 + 4\tau_1^{2}(V_x^2 + V_y^2)) \\
      &- 6e^{-(1/12)k_0^2 R_0^2} (2 k_0 R_0^2 \tau_1 V_y \cos(k_0 ((\tau_1 V_y)/3 + y_0)) \\
      &+ (-6 R_0^2 + k_0^2 R_0^4 + 8 \tau_1^{2} (V_x^2 + V_y^2)) \sin(k_0 ((\tau_1 V_y)/3 + y0))) \\
    &- 2 e^{-(3/4)k_0^2 R_0^2} (6 k_0 R_0^2 \tau_1 V_y \cos(k_0 (\tau_1 V_y + 3 y_0))+ (-6 R_0^2 + 9 k_0^2 R_0^4 \\
    &+ 8 \tau_1^{2} (V_x^2 + V_y^2)) \sin(k_0 (\tau_1 V_y + 3 y_0))))\\
    g&:=-\left(\frac{4 \alpha A \beta}{243\pi R_0^4}\right)  e^{-(3/4) k_0^2 R_0^2 - (2 \tau_1^2 (V_x^2 + V_y^2))/(3 R_0^2)} \tau_1 \cdot(3 e^{(2 k_0^2 R_0^2)/3} k_0 R_0^2 (-24 R_0^2 \\
    &+  k_0^2 R_0^4+ 4 \tau_1^2 (2 V_x^2 + 3 V_y^2)) \cos(  k_0 ((\tau_1 V_y)/3 + y_0)) \\
    &+  3 k_0 R_0^2 (-24 R_0^2 + 9 k_0^2 R_0^4 + 4 \tau_1^2(2 V_x^2 + 3 V_y^2)) \cos(k_0 (\tau_1 V_y + 3 y_0)) \\
    &+ \tau_1 V_y (-3 R_0^2 + 4 \tau_1^2 (V_x^2 + V_y^2)) (-3 e^{(3 k_0^2 R_0^2)/4} \pi \\
    &+ 12 e^{(2 k_0^2 R_0^2)/3} \sin(k_0 ((\tau_1 V_y)/3 + y_0))+ 4 \sin(k_0 (\tau_1 V_y+ 3 y_0)))).
    \end{align*}
    }
    
    For all simulations we use parameter values from Table \ref{table_params1} unless otherwise noted.
    
    \begin{table} 
         \[
         \begin{tabular}{|c|c|c|}
         \hline 
         parameter & value & description \\
         \hline 
         $R_0$ & 3 & radius of cell \\
         $\beta$ & 3 & creation of $p$ at interface \\
         $\tau_1$ & 10 & degradation of $p$ inside cell \\
         $k_0$ & 2.5 & wave number of substrate pattern \\
         $\alpha$ & 2 & advection of $\rho$ by $p$ \\
         $\eta$ & 10 & dissipation in the adhesive layer \\
         $G$ & .15 & substrate stiffness \\
         $\bar{a}_0$ & $.0025 - .25$ & linear attachment rate of adhesion sites\\
         $\bar{a}_{nl}$ & 1.5 & effective collective (nonlinear) attachment rate of adhesion sites \\
         $\bar{s} $ & 1 & local saturation of adhesion sites\\
         $d$ & 1 & detachment rate of adhesion sites \\
         $b$ & 5 & sharpness of breaking function \\
         $U_c$ & $\sqrt{5}$ & critical extension to break adhesive contacts\\
         \hline
         \end{tabular} 
         \]
         \caption{Physical parameter values. }
           \label{table_params1}
         \end{table}
 
 If we assume that $V_y=0$ we of course recover equation (14) in \cite{Zie13}.  We note that due to homogeneity of the substrate in the $x$ direction, equation $V_x = f(V_x,V_y,A,y)$ is algebraic; letting $\zeta = V_x$:
    \begin{equation}\label{dae1}
    \zeta = f(\zeta,\dot{y},y)
    \end{equation}
    \begin{equation}\label{dae2}
    \dot{y} = g(\zeta,\dot{y},y).
    \end{equation}
    Importantly \eqref{dae1}-\eqref{dae2} may not be in general solvable (or have a continuous in time solution) for all values of $\dot{y},y$. This leads to difficulty of both numerical simulations as well as prevents the use of classical linear stability analysis.
    
    \medskip

    We highlight here that the reduction described above is similar to the one conducted in \cite{Zie13} however we additionally incorporate expansions for the $V_y$ component. Moreover the equation for the effective adhesion $A_0$ is derived using the asymptotic expansion $A = A_0(t) \rho(x,y,t)+\delta A_1(x,y,t)$ which gives rise to the effective coefficients \eqref{effective_coeff}.

{\footnotesize 
\bibliographystyle{unsrt}
\bibliography{cellref}

\begin{thebibliography}{10}

\bibitem{Moh03}
R.~R. Mohan, A.~E.~K. Hutcheon, R.~Choi, J.~Hong, J.~Lee, R.~R. Mohan,
  R.~Ambr\'osio Jr., J.~D. Zieske, and S.~E. Wilson.
\newblock Apoptosis, necrosis, proliferation, and myofibroblast generation in
  the stroma following {LASIK} and {PRK}.
\newblock {\em Exp. Eye. Res.}, 76:71--87, 2003.

\bibitem{BanSte98}
J.~Banchereau and R.~M. Steinman.
\newblock Dendritic cells and the control of immunity.
\newblock {\em Nature}, 392(6673):245--252, 1998.

\bibitem{BirBir03}
C.~Birchmeier, W.~Birchmeier, E.~Gherardi, and G.~F.~V. Woude.
\newblock Met, metastasis, motility and more.
\newblock {\em Nature reviews Molecular cell biology}, 4(12):915--925, 2003.

\bibitem{Pol03}
T.~D. Pollard.
\newblock The cytoskeleton, cellular motility and the reductionist agenda.
\newblock {\em Nature}, 422(6933):741--745, 2003.

\bibitem{PolBor03}
T.~D. Pollard and G.~G. Borisy.
\newblock Cellular motility driven by assembly and disassembly of actin
  filaments.
\newblock {\em Cell}, 112(4):453--465, 2003.

\bibitem{Bar11}
E.~L. Barnhart, K.~Lee, K.~Keren, A.~Mogilner, and J.~A. Theriot.
\newblock An adhesion-dependent switch between mechanisms that determine motile
  cell shape.
\newblock {\em PLoS Biol}, 9:e1001059, 2011.

\bibitem{Abe80}
M.~Abercrombie.
\newblock The crawling movement of metazoan cells.
\newblock {\em Proc R Soc London B}, 207 (1167):129--147.

\bibitem{SheFelGal98}
M.~P. Sheetz, D.~P. Felsenfeld, and C.~G. Galbraith.
\newblock Cell migration: regulation of force on extracellular-matrix-integrin
  complexes.
\newblock {\em Trends in cell biology}, 8(2):51--54, 1998.

\bibitem{Ker08}
K.~Keren, Z.~Pincus, G.~M. Allen, E.~L. Barnhart, G.~Marriott, A.~Mogilner, and
  J.~A. Theriot.
\newblock Mechanism of shape determination in motile cells.
\newblock {\em Nature}, 453:475--480, 2008.

\bibitem{VerSviBor99}
A.~B. Verkhovsky, T.~M. Svitkina, and G.~G. Borisy.
\newblock Self-polarization and directional motility of cytoplasm.
\newblock {\em Current Biology}, 9(1):11--S1, 1999.

\bibitem{MogOdd11}
A.~Mogilner and D.~Odde.
\newblock Modeling cellular processes in 3d.
\newblock {\em Trends in cell biology}, 21(12):692--700, 2011.

\bibitem{Mog09}
A.~Mogilner.
\newblock Mathematics of cell motility: have we got its number?
\newblock {\em J. Math. Biol.}, 58:105--134, 2009.

\bibitem{CamZhaLi17}
B.~A. Camley, Y.~Zhao, B.~Li, H.~Levine, and W.-J. Rappel.
\newblock Crawling and turning in a minimal reaction-diffusion cell motility
  model: coupling cell shape and biochemistry.
\newblock {\em Physical Review E}, 95(1):012401, 2017.

\bibitem{ziebert2011model}
Falko Ziebert, Sumanth Swaminathan, and Igor~S Aranson.
\newblock Model for self-polarization and motility of keratocyte fragments.
\newblock {\em Journal of The Royal Society Interface}, page rsif20110433,
  2011.

\bibitem{RecPutTru13}
P.~Recho, T.~Putelat, and L.~Truskinovsky.
\newblock Contraction-driven cell motility.
\newblock {\em Physical review letters}, 111(10):108102, 2013.

\bibitem{RubJacMog05}
B.~Rubinstein, K.~Jacobson, and A.~Mogilner.
\newblock Multiscale two-dimensional modeling of a motile simple-shaped cell.
\newblock {\em Multiscale Modeling \& Simulation}, 3(2):413--439, 2005.

\bibitem{Sha10}
D.~Shao, W.~J. Rappel, and H.~Levine.
\newblock Computational model for cell morphodynamics.
\newblock {\em Phys. Rev. Lett.}, 105:108104, 2010.

\bibitem{Tjh15}
E.~Tjhung, A.~Tiribocchi, D.~Marenduzzo, and M.~E. Cates.
\newblock A minimal physical model captures the shapes of crawling cells.
\newblock {\em Nature communications}, 6, 2015.

\bibitem{lober2015collisions}
Jakob L{\"o}ber, Falko Ziebert, and Igor~S Aranson.
\newblock Collisions of deformable cells lead to collective migration.
\newblock {\em Scientific reports}, 5:9172, 2015.

\bibitem{ziebert2016computational}
Falko Ziebert and Igor~S Aranson.
\newblock Computational approaches to substrate-based cell motility.
\newblock {\em npj Computational Materials}, 2:16019, 2016.

\bibitem{BerPotRyb16}
L.~Berlyand, M.~Potomkin, and V.~Rybalko.
\newblock Non-uniqueness in a nonlinear sharp interface model of cell motility.
\newblock {\em Comptes Rendus Mathematique}, 354(10):986--992, 2016.

\bibitem{MizBerRybZha16}
M.~S. Mizuhara, L.~Berlyand, V.~Rybalko, and L.~Zhang.
\newblock On an evolution equation in a cell motility model.
\newblock {\em Physica D: Nonlinear Phenomena}, 318:12--25, 2016.

\bibitem{MizZha17}
M.~S. Mizuhara and P.~Zhang.
\newblock Uniqueness and traveling waves in a cell motility model.
\newblock {\em arXiv:1703.00811}, 2017.

\bibitem{RecTru16}
P.~Recho and L.~Truskinovsky.
\newblock Cell locomotion in one dimension.
\newblock In {\em Physical Models of Cell Motility}, pages 135--197. Springer,
  2016.

\bibitem{BarLeeAllTheMog15}
E.~Barnhart, K.-C. Lee, G.~M. Allen, J.~A. Theriot, and A.~Mogilner.
\newblock Balance between cell- substrate adhesion and myosin contraction
  determines the frequency of motility initiation in fish keratocytes.
\newblock {\em Proceedings of the National Academy of Sciences},
  112(16):5045--5050, 2015.

\bibitem{BarAllJulThe10}
E.~L. Barnhart, G.~M. Allen, F.~J{\"u}licher, and J.~A. Theriot.
\newblock Bipedal locomotion in crawling cells.
\newblock {\em Biophysical journal}, 98(6):933--942, 2010.

\bibitem{LobZieAra14}
J.~L{\"o}ber, F.~Ziebert, and I.~S. Aranson.
\newblock Modeling crawling cell movement on soft engineered substrates.
\newblock {\em Soft matter}, 10(9):1365--1373, 2014.

\bibitem{Zie13}
F.~Ziebert and I.~S. Aranson.
\newblock Effects of adhesion dynamics and substrate compliance on the shape
  and motility of crawling cells.
\newblock {\em PLoS ONE}, 8:e64511, 2013.

\bibitem{CsuQuiDan07}
G.~Csucs, K.~Quirin, and G.~Danuser.
\newblock Locomotion of fish epidermal keratocytes on spatially selective
  adhesion patterns.
\newblock {\em Cell motility and the cytoskeleton}, 64(11):856--867, 2007.

\bibitem{RolNakYam12}
C.~G. Rolli, H.~Nakayama, K.~Yamaguchi, J.~P. Spatz, R.~Kemkemer, and
  J.~Nakanishi.
\newblock Switchable adhesive substrates: revealing geometry dependence in
  collective cell behavior.
\newblock {\em Biomaterials}, 33(8):2409--2418, 2012.

\bibitem{CheMrkHuaWhiIng98}
C.~S. Chen, M.~Mrksich, S.~Huang, G.~M. Whitesides, and D.~E. Ingber.
\newblock Micropatterned surfaces for control of cell shape, position, and
  function.
\newblock {\em Biotechnology progress}, 14(3):356--363, 1998.

\bibitem{LoHonDemWan00}
C.-M. Lo, H.-B. Wang, M.~Dembo, and Y.-L. Wang.
\newblock Cell movement is guided by the rigidity of the substrate.
\newblock {\em Biophysical journal}, 79(1):144--152, 2000.

\bibitem{TriLeD12}
L.~Trichet, J.~Le Digabel, R.~J. Hawkins, S.~R.~K. Vedula, M.~Gupta,
  C.~Ribrault, P.~Hersen, R.~Voituriez, and B.~Ladoux.
\newblock Evidence of a large-scale mechanosensing mechanism for cellular
  adaptation to substrate stiffness.
\newblock {\em Proceedings of the National Academy of Sciences},
  109(18):6933--6938, 2012.

\bibitem{SheGar11}
A.~Shekaran and A.~J. Garcia.
\newblock Nanoscale engineering of extracellular matrix-mimetic bioadhesive
  surfaces and implants for tissue engineering.
\newblock {\em Biochimica et Biophysica Acta (BBA)-General Subjects},
  1810(3):350--360, 2011.

\bibitem{BruGouPin94}
R.~Bruinsma, M.~Goulian, and P.~Pincus.
\newblock Self-assembly of membrane junctions.
\newblock {\em Biophysical journal}, 67(2):746--750, 1994.

\bibitem{BurChrl96}
K.~Burridge and M.~Chrzanowska-Wodnicka.
\newblock Focal adhesions, contractility, and signaling.
\newblock {\em Annual review of cell and developmental biology},
  12(1):463--519, 1996.

\bibitem{SacBru02}
E.~Sackmann and R.~F. Bruinsma.
\newblock Cell adhesion as wetting transition?
\newblock {\em ChemPhysChem}, 3(3):262--269, 2002.

\bibitem{SmiSei05}
A.-S. Smith and U.~Seifert.
\newblock Effective adhesion strength of specifically bound vesicles.
\newblock {\em Physical Review E}, 71(6):061902, 2005.

\bibitem{PeyKalCohRun12}
S.~R. Peyton, Z.~I. Kalcioglu, J.~C. Cohen, A.~P. Runkle, K.~J.~Van Vliet,
  D.~A. Lauffenburger, and L.~G. Griffith.
\newblock Marrow-derived stem cell motility in 3d synthetic scaffold is
  governed by geometry along with adhesivity and stiffness.
\newblock {\em Biotechnology and bioengineering}, 108(5):1181--1193, 2011.

\bibitem{RodSch13}
L.~L. Rodriguez and I.~C. Schneider.
\newblock Directed cell migration in multi-cue environments.
\newblock {\em Integrative Biology}, 5(11):1306--1323, 2013.

\bibitem{Zie12}
F.~Ziebert, S.~Swaminathan, and I.~S. Aranson.
\newblock Model for self-polarization and motility of keratocyte fragments.
\newblock {\em J. R. Soc. Interface}, 9:1084--1092, 2012.

\end{thebibliography}
}

\end{document}